\documentclass{article}

\usepackage[numbers]{natbib} 

\usepackage[final]{neurips_2021}

\usepackage[utf8]{inputenc} 
\usepackage[T1]{fontenc}    
\usepackage{hyperref}       
\usepackage{url}            
\usepackage{booktabs}       
\usepackage{amsfonts}       
\usepackage{nicefrac}       
\usepackage{booktabs}
\usepackage{microtype}      
\usepackage{xcolor}         
\usepackage{graphicx}
\newcommand{\RR}{I\!\!R} 

\newcommand\red[1]{{\color{black}{#1}}}
\usepackage{scalefnt}

\usepackage{amsmath}
\usepackage{amsmath}
\usepackage{amssymb}
\usepackage{physics}
\usepackage{mathtools}
\usepackage{verbatim}
\usepackage{xr}
\usepackage{bbold}
\usepackage{multirow}
\usepackage{tabu}
\usepackage{placeins}

\setcitestyle{square}

\title{Towards efficient end-to-end speech recognition \\with biologically-inspired neural networks}

\newcommand{\figTwo}[1]{
\begin{figure*}[#1]
\centering
\includegraphics[width=1.\columnwidth]{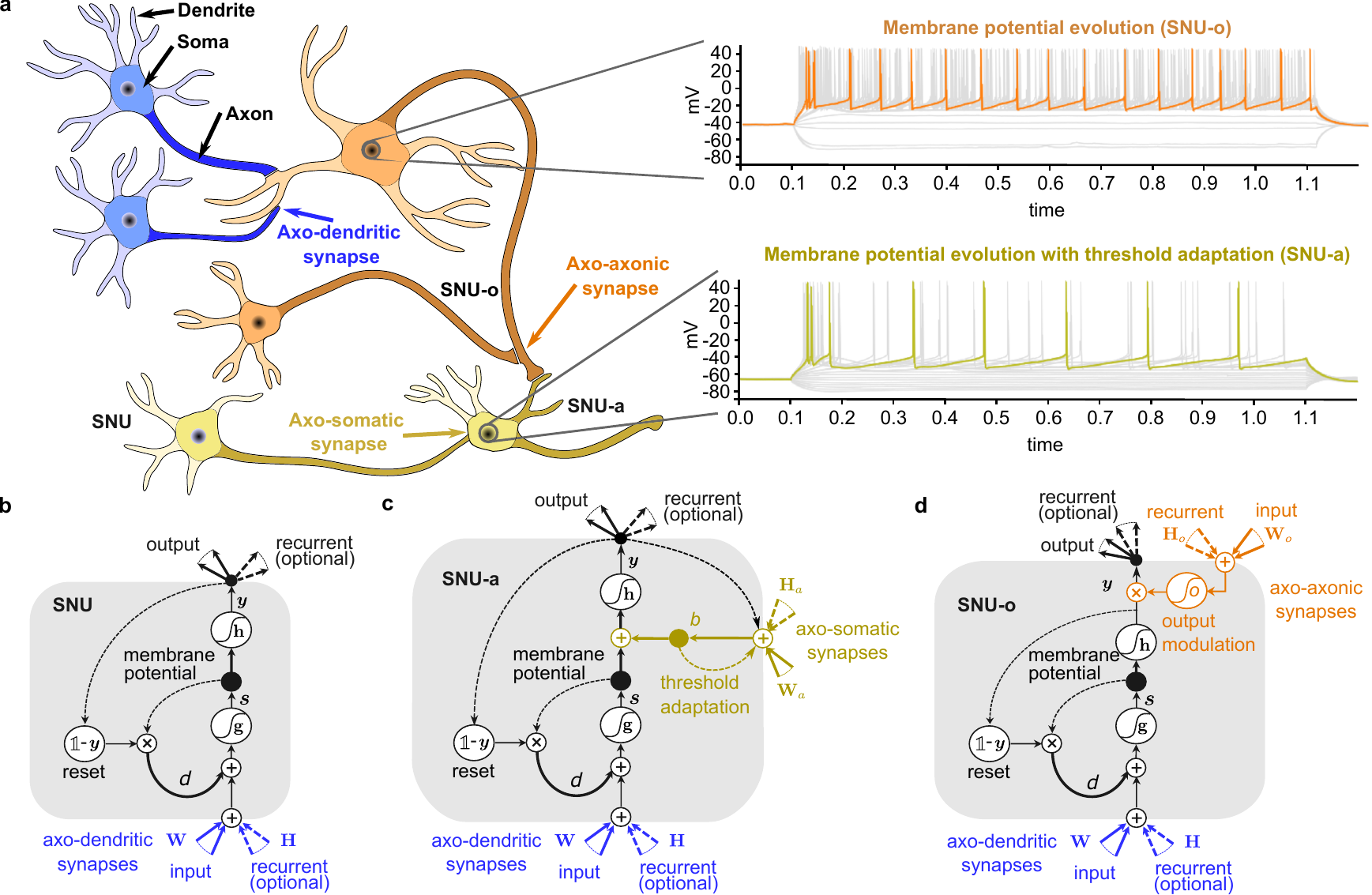}
\caption{\textbf{Illustration of different biological synapse and neuron types and their realization in simulations}. \textbf{a} A neural network example composed of three neuron and three synapse types along with recordings of membrane potentials from the Human Brain Atlas~\cite{Allen2010}. \textbf{b} Visualization of SNU modelling the LIF dynamics using solely axo-dendritic synapses. \textbf{c} Visualization of the SNU-a modelling the adaptive threshold behavior using axo-somatic synapses, illustrated with yellow color in \textbf{a}. \textbf{d} Visualization of the SNU-o using axo-axonic synapses, illustrated in orange in \textbf{a}.}
\label{fig:Figure2}
\end{figure*}
}

\newcommand{\figThree}[1]{
\begin{figure*}[#1]
\centering
\includegraphics[width=1.\columnwidth]{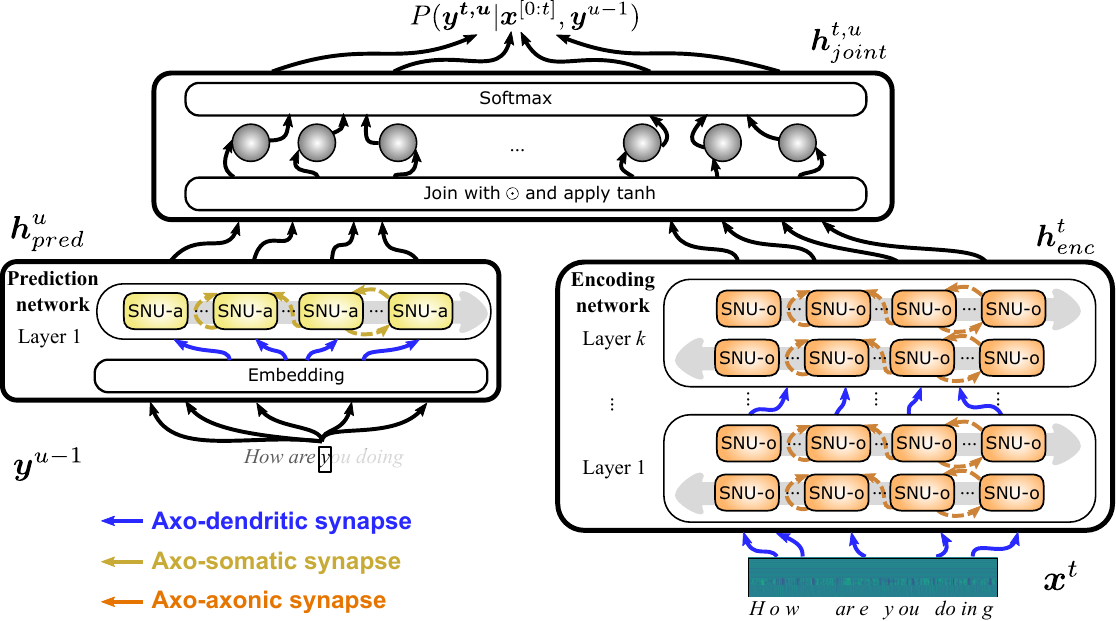}
\caption{\textbf{Illustration of the RNN transducer architecture}. An input vector $\boldsymbol{x}^t$, containing MFCC features of a speech signal, is processed by the encoding network, here represented with sSNU-o units. The prediction network, corresponding to a language model, here represented with sSNU-a units, enhances the predictions based on the past output sequence of the RNN-T, i.e. $\boldsymbol{y}^{u-1}$. The joint network forms the final predictions based on the outputs from both subnetworks.
}
\label{fig:Figure3}
\end{figure*}
}

\newcommand{\figSix}[1]{
\begin{figure*}[#1]
\centering
\includegraphics[width=1.\columnwidth]{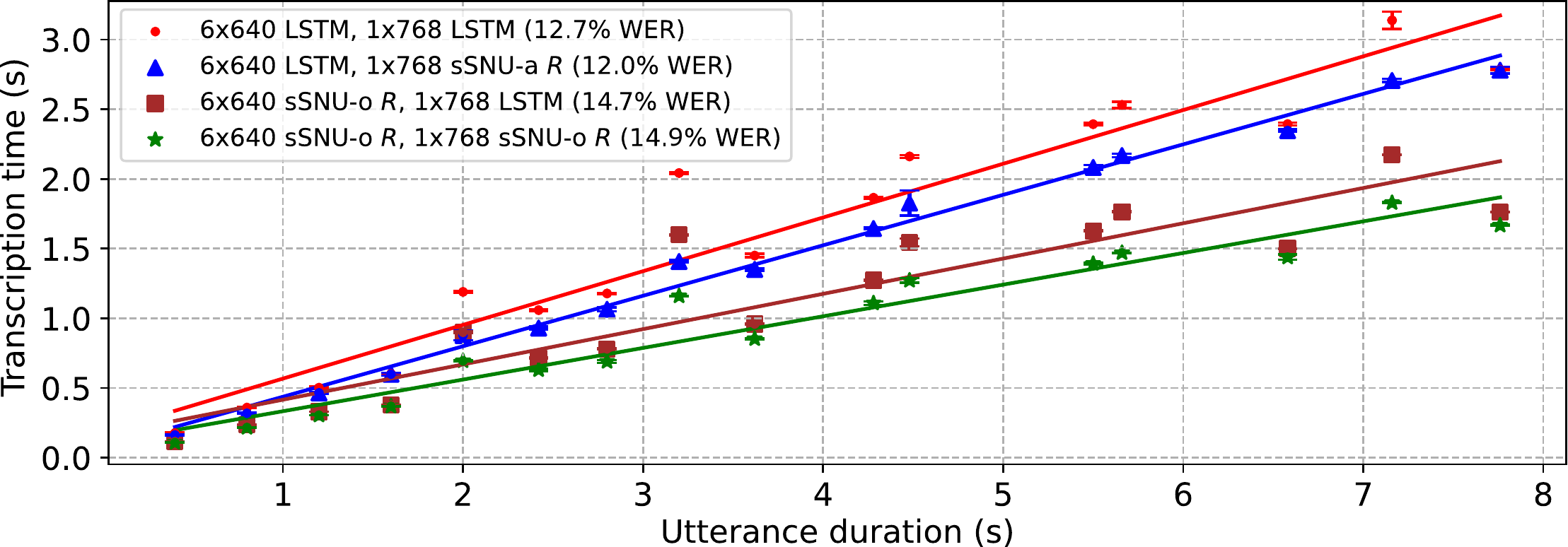}
\caption{\textbf{Comparison of the transcription time of various architectures.} The data points represent the mean time taken to decode utterances of different lengths. The markers indicate the various architectures used and the error bars depict the standard deviation over 10 executions.}
\label{fig:Figure4}
\end{figure*}
}

\newcommand{\toptitlebar}{
  \hrule height 4px
  \vskip 0.25in
  \vskip -\parskip%
}
\newcommand{\bottomtitlebar}{
  \vskip 0.29in
  \vskip -\parskip
  \hrule height 1px
  \vskip 0.09in%
}

\newcommand{\supTitle}{
\vbox{
    \vskip 0.3in
    \toptitlebar
    \centering
    {\LARGE\bf Supplementary material\par}
    \bottomtitlebar
}
}

\author{%
  Thomas Bohnstingl\thanks{Correspondence to boh@zurich.ibm.com \newline \hspace*{1.4em}$\dagger$ Currently with \red{Axelera AI, }Zurich}\\
  IBM Research, Zurich\\
  Graz University of Technology\\
  \And
  Ayush Garg\\
  IBM Research, Zurich\\
  ETH Zurich\\
  \And
  Stanisław Woźniak\\
  IBM Research, Zurich\\
  \And
  George Saon\\
  IBM Research AI, Yorktown Heights\\
  \And
  Evangelos Eleftheriou$\dagger$\\
  IBM Research, Zurich\\
  \And
  Angeliki Pantazi\\
  IBM Research, Zurich\\
}

\begin{document}

\maketitle

\begin{abstract}
   Automatic speech recognition (ASR) is a capability which enables a program to process human speech into a written form. Recent developments in artificial intelligence (AI) have led to high-accuracy ASR systems based on deep neural networks, such as the recurrent neural network transducer (RNN-T). However, the core components and the performed operations of these approaches depart from the powerful biological counterpart, i.e., the human brain. On the other hand, the current developments in biologically-inspired ASR models, based on spiking neural networks (SNNs), lag behind in terms of accuracy and focus primarily on small scale applications. In this work, we revisit the incorporation of biologically-plausible models into deep learning and we substantially enhance their capabilities, by taking inspiration from the diverse neural and synaptic dynamics found in the brain. In particular, we introduce neural connectivity concepts emulating the axo-somatic and the axo-axonic synapses. Based on this, we propose novel deep learning units with enriched neuro-synaptic dynamics and integrate them into the RNN-T architecture. We demonstrate for the first time, that a biologically realistic implementation of a large-scale ASR model can yield competitive performance levels compared to the existing deep learning models. Specifically, we show that such an implementation bears several advantages, such as a reduced computational cost and a lower latency, which are critical for speech recognition applications.
\end{abstract}

\section{Introduction}
\label{sec:Introduction}
Speech is one of the most natural ways for human beings to communicate and to interact, which led several research groups into developing speech recognition applications using computer systems. However, even if the origin of the sound signal is the same for the humans and the computers, the transformation, the feature extraction and the further processing differ significantly. Particularly, the later often utilizes mathematical abstractions, such as the Hidden Markov Model (HMM) or domain-specific models~\cite{Baker1975, Gales2007, Benzeghiba2007Oct, Yu2015}.
More recently, AI researchers have increasingly resorted to deep learning approaches ~\cite{Graves2004, juang2005, Xiong2017Mar, Nguyen2020Oct}. Especially three kinds of architectures have been widely adopted in the literature~\cite{Wang2019Aug, ComparisonOfASR}: architectures based solely on recurrent neural networks (RNN), such as the RNN transducer~\cite{Graves2012Nov}, architectures based on RNNs with the attention mechanism, such as the listen attend and spell model (LAS)~\cite{Chan2016}, and recently also transformer-based models, such as the conformer~\cite{gulati20_interspeech}. While the transformer-based and attention-based models potentially provide higher accuracy, \red{they require the entire audio signal to compute the transcript and hence have a high latency. In contrast, t}he RNN-T architecture\red{ can produce a transcript while receiving the audio signal and thus }exhibits a lower latency, allows for real-time transcription and is even commercially deployed, for example in mobile devices~\cite{He2018Nov}. Recently, researchers have also tried to merge the RNN-T and the LAS model in order to leverage the benefits of both~\cite{Sainath2019, Hu2020Mar, Hu2021}.

Despite their great successes, all the aforementioned models take inspiration from biology only remotely and depart considerably from the human speech recognition system. Researchers working on biologically plausible spiking neural networks (SNNs), have used the leaky integrate-and-fire (LIF) neurons for speech recognition~\cite{Dennis2013, Wu2018, Dong2019May}. However, there are several limitations of these works. For example, many of them use simpler network architectures compared to the ones from the ML domain. Moreover, often only parts of the network architecture employ biologically-inspired units and the learning algorithms used are inferior to error-backpropagation~\cite{Gutig2006Mar, Shi2021Mar}. Thus, the performance of those approaches lags behind their ML-based counterparts in terms of accuracy. 

In this paper, we address these limitations by proposing an RNN-T architecture that incorporates biologically-inspired neural units. Specifically, we leverage the diverse neuron and synapse types observed in the brain and propose novel extensions of the LIF model with much richer dynamics. We build upon the spiking neural units (SNUs)~\cite{Wozniak2020Jun}, which allows us to leverage the advanced training capabilities from the ML domain~\cite{Kingma2014Dec,Loshchilov2017Nov} \red{that} are essential for speech recognition~\cite{Saon2021Mar}. We demonstrate that a state-of-the-art network incorporating biologically-inspired units can yield competitive performance levels in a large-scale speech recognition application, while significantly reducing the computational cost as well as the latency. In particular, our key contributions are:

\begin{itemize}
    \item a novel neural connectivity concept emulating the axo-somatic synapses that enhances the threshold adaptation in biologically-inspired neurons,
    \item a novel neural connectivity concept emulating the axo-axonic synapses that modulates the output of biologically-inspired neurons,
    \item a biologically-inspired RNN-T architecture with significantly reduced computational cost and increased throughput, which still provides a competitive performance to the LSTM-based architecture.
\end{itemize}

\section{Biologically-inspired models for deep learning}
\label{sec:bioPlausibleNN}
Typically, architectures of biologically-inspired neurons consider only axo-dendritic synapses, in which the output from the pre-synaptic neuron travelling down the axon is modulated by the synaptic weight and arrives at the post-synaptic dendrite. However, neural networks in the brain exhibit a much more complex connectivity structure~\cite{Howard2005Jun, bear2020neuroscience}. In this work, we leverage this diversity and investigate two additional types of synapses, namely the axo-somatic and the axo-axonic synapses, and propose novel biologically-inspired models with richer dynamics. Figure~\ref{fig:Figure2}a shows an illustration of such neuron models and their connectivity via the various types of synapses.

\figTwo{!t}

The simplest biologically-inspired model is based on the LIF dynamics \red{which employs axo-dendritic synapses only, see dark blue color in Figure~\ref{fig:Figure2}a}. \red{It} can be abstracted into the form of a recurrent unit using the SNU framework~\cite{Wozniak2018Dec}. Figure~\ref{fig:Figure2}b shows the basic configuration of the SNU. A layer of $n$ SNUs receiving $m$ inputs is governed by the following equations:
\begin{align}
\boldsymbol{s}^t &= \mathbf{g}(\mathbf{W} \boldsymbol{x}^t + \mathbf{H} \boldsymbol{y}^{t-1} + d \cdot \boldsymbol{s}^{t-1} \odot (\mathbb{1} - \boldsymbol{y}^{t-1})) \label{eq:SNUState},\\
\boldsymbol{y}^t &= \mathbf{h}(\boldsymbol{s}^t + \mathbf{b}), \label{eq:SNUOutput}
\end{align}
where $\boldsymbol{x}^t \in \RR^{m}$ represents the inputs at time step $t$, $\boldsymbol{s}^t \in \RR^{n}$ represents the membrane potentials\red{, the hidden states, }
, $\boldsymbol{y}^{t} \in \RR^{n}$ represents the outputs
, $d \in \RR$ represents the constant decay of the membrane potential, $\mathbf{b} \in \RR^{n}$ represents the trainable firing threshold, $\mathbf{W} \in \RR^{n\,\times\,m}$ and $\mathbf{H} \in \RR^{n\,\times\,n}$ denote the trainable input and the recurrent weight matrices. Note that \red{although the SNU shares similarities with an RNN, e.g., the term $\mathbf{W} \boldsymbol{x}^t + \mathbf{H} \boldsymbol{y}^{t-1}$ and the activation function $\mathbf{h}$, it has a richer dynamics, including a state decay and an internal state reset mechanism, i.e., $d \cdot \boldsymbol{s}^{t-1} \odot (\mathbb{1} - \boldsymbol{y}^{t-1})$.
}
As described in~\cite{Wozniak2020Jun}, the SNU can operate in principle in two different modes, one in which 
it yields continuous outputs (called sSNU), i.e. $\mathbf{h}=\boldsymbol{\sigma}$, where $\boldsymbol{\sigma}$ denotes the sigmoid function, and one in which
it emits discrete spikes, i.e. $\mathbf{h}=\boldsymbol{\Theta}$, where $\boldsymbol{\Theta}$ indicates the Heaviside function. In this work, we focus on the dynamics of the neuronal models, in particular on the different synapse types, and thus mainly consider the sSNU variants.

One well-known property of neurons in the human brain is that they can adapt their firing threshold across a wide variety of timescales~\cite{Allen2010, Bellec2018b, Wozniak2018Dec}. In biology, there are complex mechanisms influencing the firing threshold. In particular, it can be increased or decreased following the neuron's own dynamics, but also based on the activity of the other neurons~\cite{Fontaine2014Apr, Huang2016Jun}. We propose a novel adaptive SNU (SNU-a) that enhances the threshold adaptivity dynamics by emulating the axo-somatic synapses, indicated with yellow color in Figure~\ref{fig:Figure2}a. A layer of $n$ SNU-a units is governed by
\begin{align}
\boldsymbol{s}^t &= \mathbf{g}(\mathbf{W} \boldsymbol{x}^t + \mathbf{H} \boldsymbol{y}^{t-1} + d \cdot \boldsymbol{s}^{t-1} \odot (\mathbb{1} - \boldsymbol{y}^{t-1})) \label{eq:SNUAState},\\
\boldsymbol{b}^t &= \boldsymbol{\rho} \odot \boldsymbol{b}^{t-1} + \left(1-\boldsymbol{\rho}\right) \odot \left(\mathbf{W}_a \boldsymbol{x}^t +  \mathbf{H}_a \boldsymbol{y}^{t-1}\right)\label{eq:SNUAThreshold},\\
\boldsymbol{y}^t &= \mathbf{h}(\boldsymbol{s}^t + \beta \boldsymbol{b}^t + \mathbf{b}_0), \label{eq:SNUAOutput}
\end{align}
where $\boldsymbol{b}_0 \in \RR^{n}$ represents the trainable baseline threshold, $\beta \in \RR$ represents a constant scaling factor, $\boldsymbol{b}^t \in \RR^{n}$ represents the state of the threshold at time $t$, $\boldsymbol{\rho} \in \RR^{n}$ represents the constant decay of the threshold, $\mathbf{W}_a \in \RR^{n\,\times\,m}$ and $\mathbf{H}_a \in \RR^{n\,\times\,n}$ denote the trainable input and recurrent weight matrices influencing the threshold via axo-somatic synapses. Figure~\ref{fig:Figure2}c shows an illustration of the SNU-a and the lower right part of Figure~\ref{fig:Figure2}a shows an example of the membrane potential evolution.

Another type of neural connectivity in the human brain is via axo-axonic synapses, indicated with orange color in Figure~\ref{fig:Figure2}a. These synapses mediate the release of neurotransmitters from the pre-synaptic to the post-synaptic neuron. Interneurons (INs), e.g. Chandelier INs, that connect via axo-axonic synapses are found in the brain and often act as inhibitors for the post-synaptic neuron~\cite{tremblay2016gabaergic, Fishell2011Jun}. However, it has also been found that such neurons may be excitatory as well~\cite{Fishell2011Jun}. We incorporate the axo-axonic synapses into the SNU framework and propose SNU-o units, with the following equations
\begin{align}
\boldsymbol{s}^t &= \mathbf{g}(\mathbf{W} \boldsymbol{x}^t + \mathbf{H} \boldsymbol{y}^{t-1} + d \cdot \boldsymbol{s}^{t-1} \odot (\mathbb{1} - \boldsymbol{\tilde{y}}^{t-1})),\label{eq:SNUOState}\\
\boldsymbol{\tilde{y}}^t &= \mathbf{h}(\boldsymbol{s}^t + \boldsymbol{b}^t),\\
\boldsymbol{y}^t &= \boldsymbol{\tilde{y}}^t \odot \mathbf{o}\left(\mathbf{W}_o \boldsymbol{x}^t + \mathbf{H}_o \boldsymbol{y}^{t-1} + \boldsymbol{b}_o^t\right)\label{eq:SNUOOutput}
\end{align}
where $\boldsymbol{\tilde{y}^t}$ represents the unmodulated output of the neuron driving the neural reset, $\boldsymbol{y}^t$ represents the modulated output of the neuron propagating to other connected neurons\red{. The trainable parameters} $\mathbf{W}_o \in \RR^{n\,\times\,m}$ and $\mathbf{H}_o \in \RR^{n\,\times\,n}$ denote the input and recurrent weight matrices and $\boldsymbol{b}_o^t \in \RR^{n}$ denotes the trainable bias term that all three modulate the neuronal output via axo-axonic connections. Note that in our simulations we use the sigmoid function as the activation for the output modulation in Equation~\ref{eq:SNUOOutput}, i.e., $\mathbf{o}=\boldsymbol{\sigma}$, to mimic the inhibitory character of these synapses. However, by using a different activation function $\mathbf{o}$, the outputs can also be modulated in an excitatory manner. 
Figure~\ref{fig:Figure2}d shows an illustration of the SNU-o with the output modulation and the upper right part of Figure~\ref{fig:Figure2}a shows the connectivity motif along with an example of the membrane potential evolution.

As these novel units provide different dynamics than those of the predominantly used LSTM units, we include a more detailed comparison of their dynamics in the supplementary material~1. Note that although the subsequently presented models use only a single type of synapses per neuron, multiple synapse types can be combined, resulting in further enriched neural dynamics.

\section{RNN-T with biologically-inspired dynamics}
\label{sec:BioRNNT}
The network architecture of RNN-T, illustrated in Figure~\ref{fig:Figure3}, consists of two main subnetworks, the encoding network and the prediction network, whose outputs are combined to ultimately form the speech transcript. More details of the network architecture are presented in supplementary material~2.

\figThree{b!}

In this work, we redesign the RNN-T architecture, using the novel units introduced in Section~\ref{sec:bioPlausibleNN}. In order to do so, we follow a three-step approach. First, the units are introduced into the prediction network, replacing the LSTMs, while the encoding network remains composed of LSTM units. In a second step, we incorporate the sSNU variants into the encoding network only, i.e., the prediction network remains composed of LSTMs. Finally, the sSNU variants are integrated into both network parts and thus the full RNN-T architecture is composed of sSNU units. The encoding network carries out a different task than the prediction network and hence different units might be better suited. Note that a composition of diverse units and synapse types is also observed in the brain.

\red{S}peech recognition requires a large amount of compute resources. Thus, reducing the computational cost is of paramount importance. Our proposed units not only reflect the biological inspirations, but additionally are simpler in terms of the number of gates and parameters than LSTMs\red{. Hence, they }provide the potential to drastically reduce the computational cost and the \red{transcription time}.

\section{Experiments and results}
\label{sec:Results}
The simulations were performed using the Switchboard speech corpus comprising roughly 300 hours of English telephone conversations.
In order to be comparable to the state-of-the-art literature, we closely followed the preprocessing, the evaluation, as well as the network architecture setup presented in~\cite{Saon2021Mar}.
An LSTM-based version of this RNN-T architecture achieves a \red{word error rate (WER)} of $12.7\,\%$, which we consider as our baseline. We investigated various configurations of the sSNU variants, listed with their abbreviations and trainable parameters in \red{Table~\ref{tab:RNNsuffix}}.
In addition, supplementary material~3 provides \red{details about the dataset and its preprocessing, as well as }hyperparameter settings.


\begin{table}[]
    \caption{RNN acronyms used in the result tables and details on the trainable parameters.}
    \label{tab:RNNsuffix}
    \centering
    \begin{tabular}{l|l|l|l|l|l}
         \hline
         RNN \textit{Suffix} & Comment & Thr. & \shortstack{Axo-\\dendritic} & \shortstack{Axo-\\somatic} & \multicolumn{1}{|c}{\shortstack{Axo-\\axonic}} \\
         \hline
         sSNU & Feedforward &  $\mathbf{b}$ & $\mathbf{W}$ &  \\
         sSNU \textit{R} & Recurrent &  $\mathbf{b}$ &  $\mathbf{W, H}$ & \\
         sSNU-a &  Adaptive thr. feedforward & $\mathbf{b_0}$ & $\mathbf{W}$ & & \\
         sSNU-a \textit{R} & Adaptive thr. recurrent & $\mathbf{b_0}$ & $\mathbf{W, H}$ & & \\
         sSNU-a \textit{Ra} & Adaptive thr. axo-somatic recurrent & $\mathbf{b_0}$ & $\mathbf{W, H}$ & $\mathbf{H_a}$ & \\
         sSNU-o &  Output modulating feedforward &  $\mathbf{b}$ & $\mathbf{W}$ & & $\mathbf{W_o, b_o}$ \\
         sSNU-o \textit{R} & Output modulating recurrent &   $\mathbf{b}$ & $\mathbf{W, H}$ & & $\mathbf{W_o, H_o, b_o}$ \\
         \hline
    \end{tabular}
\end{table}

\begin{table}[b!]
  \caption{Performance comparison of the RNN-T network
  \red{for different configurations of the prediction and the encoding network}
  }
  \label{tab:perfNW}
  \centering
  \begin{tabu}{l|l|c|rr|rr|rr}
    \hline
    Enc. RNN & Pred. RNN & WER (\%) & \# Param. & (\%) & \# Multipl. & (\%) & $t_{inf}$(s) & (\%) \\
    \tabucline[2pt]{-}
    \multirow{7}{*}{LSTM} & LSTM & 12.7 & 2.39M & 100 & 2.39M & 100 & 2.78 & 100 \\
    &sSNU & 15.1 & 8.45k & < 1 & 9.2k & < 1 & 2.71 & 97\\
    &sSNU \textit{R} & 12.4 & 0.60M & 25 & 0.60M & 25 & 2.76 & 99 \\
    &sSNU-a & 12.1 & 8.45k & < 1 & 11.52k & < 1 & 2.73 & 98\\
    &sSNU-a \textit{R} & 12.0 & 0.60M & 25 & 0.60M & 25 & 2.78 & 100\\ 
    &sSNU-o & 12.6 & 16.90k & < 1 & 17.66k & <1 & 2.75 & 98\\
    &sSNU-o \textit{R} & 12.4 & 1.20M & 50 & 1.20M & 50 & 2.76 & 99\\
    \tabucline[2pt]{-}
    LSTM & \multirow{4}{*}{LSTM} & 12.7 & 54.20M & 100 & 54.20M & 100 &  2.78 & 100\\
    sSNU-a \textit{Ra} & & 25.2 & 18.47M & 34 & 18.\red{50}M & 34 & 1.23 & 44\\
    sSNU-o & & 23.2 & 17.27M & 32 & 17.28M & 32 & 1.22 & 44\\
    sSNU-o \textit{R} & & 14.7 & 27.10M & 50 & 27.11M & 50 & 1.76 & 63\\
    \tabucline[2pt]{-}
    LSTM & LSTM & 12.7 & 5\red{6.59}M & 100 & 5\red{6.58}M & 100 & 2.78 & 100\\
    sSNU-o \textit{R} & sSNU-a \textit{R} & 16.0 & 27.70M & 49 & 27.71M & 49 & 1.64 & 59\\
    sSNU-o \textit{R} & sSNU-o \textit{R} & 14.9 & 28.30M & 50 & 28.30M & 50 & 1.66 & 60 \\
    \hline
  \end{tabu}
\end{table}
\red{\textbf{Comparison of the word error rate.}} The first part of Table~\ref{tab:perfNW} summarizes the results of the RNN-T architecture, where the sSNU variants were integrated into the prediction network. Several configurations, e.g., sSNU \textit{R} (12.4\% WER), sSNU-a \textit{R} (12.0\% WER) and sSNU-o \textit{R} (12.4\% WER), outperform the LSTM-based variant. 
\red{In a second step,} we evaluated the performance of the encoding network composed of sSNU-a and sSNU-o units. The best performance is achieved with sSNU-o \textit{R} with 
14.7\% WER compared to 12.7\% WER of the LSTM baseline. 
Finally, the sSNU units are integrated into both subnetworks so that the RNN-T is solely based on sSNUs, see the last part of Table~\ref{tab:perfNW}. Consistent with the prior two cases, the RNN-T architecture achieved competitive performance with 16.0\% WER and 14.9\% WER.
\red{\textbf{Evaluation of the inference time.}}
The inference time is a critical metric for speech recognition that depends on the model architecture and the efficiency of the neural units. In our case, this efficiency is largely dominated by the number of multiplications, including vector-matrix and scalar multiplications, and the number of activations that the units require to compute. Thus, we accumulated the total time, running on CPUs, for all computations involved during the inference of a single utterance. The last three columns in Table~\ref{tab:perfNW} show the number of parameters, the number of multiplications as well as the average time taken for the greedy decoding of an utterance with $T=388$ input frames (\textasciitilde7.8s of audio input). 

\red{When incorporating the sSNUs into the prediction network, the number of multiplications of this subnetwork is reduced substantially, see the first part of Table~\ref{tab:perfNW}. However, this is not fully reflected in the transcription time, because this subnetwork contributes only a small fraction to the total computational cost of the RNN-T.}
As one can see in the second part of Table~\ref{tab:perfNW}, when the LSTM units of the encoding network are replaced with sSNUs, 
the total number of parameters is reduced by more than 50\%
and this results in a more than \textasciitilde40\% reduction of the inference time. A similar trend can be observed in the final part of Table~\ref{tab:perfNW}, where the RNN-T is solely composed of sSNUs. In this case, the RNN-T achieves a competitive WER, albeit using 50\% fewer parameters and experiencing a \textasciitilde$40$\% reduce inference time.

Figure~\ref{fig:Figure4} depicts a comparison of the transcription time to decode utterances of various lengths for selected models. To ensure reproducibility and consistency, the examples were chosen such that the transcripts of all models \red{were identical,} and the timing results were averaged over 10 repetitions. A general observation is that the time taken to transcribe utterances increases proportionally to the utterance length. The RNN-T architecture composed of LSTM units is indicated with red dots and is the slowest among the evaluated models. Depending on the sSNU variant used, the inference time can be reduced. For example, the RNN-T using sSNU-o units for the encoding and sSNU-o units for the prediction network (6x640 sSNU-o \textit{R}, 1x768 sSNU-o \textit{R}), has an approximately $40\%$ reduced inference time.

\figSix{t!}

\section{Conclusions}
\label{sec:Conclusion}
In this work we combine insights from neuroscience with a state-of-the-art network architecture from machine learning and apply it to the task of speech recognition. We resort to the diverse types of synapses present in the brain and enhance the dynamics of the commonly used LIF neuron model with a threshold adaptation mechanism based on axo-somatic synapses as well as with an output modulating mechanism based on the axo-axonic synapses. We successively integrate those novel units into the RNN-T architecture and \red{not only }demonstrate that end-to-end speech recognition \red{with biologically-inspired units is possible, but also that they} achieve a competitive performance level of 14.9\% WER, compared to 12.7\% WER of the LSTM-based RNN-T. \red{Moreover}, the introduced units are simpler than LSTMs, so that the computational costs, as well as the inference time can be reduced by 50\% and 40\% respectively. Finally, it is worth mentioning that biology provides an abundance of mechanisms, which could further enrich the dynamics of neurons, and that are not yet covered in practical neural networks. Our simulation results indicate that such effects can potentially bolster the performance of biologically-inspired neural units and are therefore an essential step going forward.




\supTitle


\setcounter{section}{0}
\setcounter{equation}{0}
\setcounter{table}{0}

\section{Comparison of biologically-inspired units to LSTMs}
As discussed in Section~1 of the main paper, LSTM units are commonly used in state-of-the-art machine learning networks for speech recognition. They have three distinct gates which mediate the input to the units, the decay of the internal state as well as the output of the units. In particular, a layer of $n$ LSTM units with $m$ inputs is governed by the following equations
\begin{align}
\mathbf{i}^t &= \boldsymbol{\sigma}(\mathbf{W}_i \boldsymbol{x}^t + \mathbf{H}_i \boldsymbol{y}^{t-1}+\mathbf{b}_i),\\
\boldsymbol{c}^t &= \boldsymbol{\sigma}(\mathbf{W}_c \boldsymbol{x}^t + \mathbf{H}_c \boldsymbol{y}^{t-1}+\mathbf{b}_c)\label{eq:LSTMOutputGate}\\
\mathbf{f}^t &= \boldsymbol{\sigma}(\mathbf{W}_f \boldsymbol{y}^t + \mathbf{H}_f \boldsymbol{y}^{t-1}+\mathbf{b}_f),\\
\boldsymbol{s}^t &= \mathbf{f}^t \odot \boldsymbol{s}^{t-1} +\mathbf{i}^t \odot \mathbf{tanh}(\mathbf{W}_s \boldsymbol{y}^t + \mathbf{H}_s \boldsymbol{y}^{t-1}+\mathbf{b}_s),\\
\boldsymbol{y}^t &= \mathbf{c}^t \odot \mathbf{tanh}(s^t)\label{eq:LSTMOutput}.
\end{align}

We found empirically that 
the standard LIF dynamics lacks such functionalities and thus has intrinsic limitations in tackling ASR in comparison with LSTMs. For example, the dynamics of the output gate in LSTMs is governed by Equation~\ref{eq:LSTMOutputGate},
where $\mathbf{W}_c \in \RR^{n\,\times\,m}$, $\mathbf{H}_c \in \RR^{n\,\times\,n}$, $\mathbf{b}_c \in \RR^{n}$ are trainable parameters. The novel variants of SNUs introduced in Section~2 of the main paper, exploit additional dynamics 
beyond the common LIF model, achieved via the axo-somatic and axo-axonic synapses. In particular, the threshold adaptation of the SNU-a, as represented in Equation~4 of the main paper, controls the output of the neuron by increasing or decreasing the firing threshold.
By comparing the output mechanisms of the LSTM to the SNU-a
\begin{align*}
\textrm{Output gate LSTM:}&\,\,\,\,\boldsymbol{\sigma}(\mathbf{W}_c \boldsymbol{x}^t + \mathbf{H}_c \boldsymbol{y}^{t-1}+\mathbf{b}_c),\\
\textrm{Adaptive threshold SNU-a:}&\,\,\,\,\beta \boldsymbol{b}^t,
\end{align*}
with
\begin{align*}
\boldsymbol{b}^t &= \boldsymbol{\rho} \odot \boldsymbol{b}^{t-1} + \left(1-\boldsymbol{\rho}\right) \odot \left(\mathbf{W}_a \boldsymbol{x}^t + \mathbf{H}_a \boldsymbol{y}^{t-1}\right),
\end{align*}
one can see that the adaptive threshold mechanism is different from the LSTM gate. The threshold adaptation presents a low-pass filter, with a constant decay of $\boldsymbol{\rho}$, of the input activity and the recurrent activity multiplied with a trainable matrix. Moreover, 
its relation to the neuronal output is additive, whereas for the output gates in LSTM units it is multiplicative,
see Equations~5 of the main paper and Equation~\ref{eq:LSTMOutput} of these notes.

In contrast, the output modulating mechanism of the SNU-o, mimicking the axo-axonic synapses, is more closely related to the LSTM output gate. This becomes apparent when comparing the relevant equations for the LSTM units and SNU-o units:
\begin{align*}
\textrm{Output gate LSTM:}\,\,\,\,&\boldsymbol{\sigma}(\mathbf{W}_c \boldsymbol{x}^t + \mathbf{H}_c \boldsymbol{y}^{t-1}+\mathbf{b}_c),\\
\textrm{Ouput modulation SNU-o:}\,\,\,\,&\boldsymbol{\sigma}\left(\mathbf{W}_o \boldsymbol{x}^t + \mathbf{H}_o \boldsymbol{y}^{t-1} + \boldsymbol{b}_o^t\right).
\end{align*}
It is important to mention that although the output modulating mechanism of the SNU-o resembles closer the behavior of the LSTM output gate, the former mechanism has a different background and is inspired from the existence of different synapse types in the human brain.

\section{Details of the RNN-T architecture}
As described in the main paper, we utilize a state-of-the-art RNN-T network and redesign it incorporating biologically-inspired units. Broadly speaking, the RNN-T consists of two main network components leveraging RNNs, the encoding network and the prediction network.

The encoding network is responsible for the feature encoding and processes the MFCCs, denoted with $\boldsymbol{x}^t={\boldsymbol{x}^0,\boldsymbol{x}^1,\dots,\boldsymbol{x}^{T-1}}\in\RR^{n_{MFCC}}$, where $n_{MFCC}$ is the number of mel frequency cepstral coefficients and $T$ is the length of the input sequence. The encoding network uses $k$ bidirectional layers, we use $k=6$ in our simulations. 
The second part, the prediction network, acts as a language model and processes the output produced thus far by the RNN-T without the blank symbols, i.e., $\boldsymbol{y}^u={\boldsymbol{y}^0,\boldsymbol{y}^1,\dots,\boldsymbol{y}^{U}}\in\RR^{(n_{voc}+1)}$. Here $n_{voc}$ is the size of the vocabulary which in our case consists of $45$ individual characters, $U$ is the length of the final output sequence, which might be different from the input sequence length $T$ and the initial input to the prediction network $\boldsymbol{y}^0$ is always the blank symbol. Note that the prediction network is composed of an embedding layer followed by a layer of recurrent neural units.
The outputs of the encoding network, the acoustic embedding, $\boldsymbol{h}^t_{enc}\in\RR^{n_{enc}}$, where $n_{enc}$ is the number of units in the last layer of the encoder, and the output of the prediction network, the prediction vector $\boldsymbol{h}^u_{pred}\in\RR^{n_{pred}}$ where $n_{pred}$ is the number of units in the last layer of the prediction network, are expanded, combined together via a Hadamard product and then further processed by a $\mathbf{tanh}$ activation and softmax operation. The final result, $\boldsymbol{h}^{t,u}_{joint}\in\RR^{T\,\times\,U\,\times\,(n_{voc}+1)}$, is then used to compute the output distribution and in turn the most probable input-output alignment using the Forward-Backward algorithm as proposed in~\cite{Graves2012Nov}. Note that the output of the joint network may contain a special blank symbol that allows for alignment of the speech signal with the transcript. This symbol gets removed from the final prediction of the RNN-T network.

\section{Data preprocessing and simulation details}
In our work we investigate the Switchboard speech corpus, which is a widely adopted dataset of roughly 300 hours of English two-sided telephone conversations on predefined topics. In particular, the dataset contains speech from a total of 543 speakers from different areas of the United States and is licensed under the Linguistic Data Consortium~\cite{BibEntry2021May}.

Initially, four data augmentations are applied to the original dataset in which the speed as well as the tempo of the conversation are increased and decrease by a value of $1.1$ and $0.9$, respectively. Then, a 40-dimensional MFCC vector is extracted every $10$\,ms and extended with the first and second order derivatives, yielding a 120-dimensional vector. Next, a time reduction technique is applied that involves stacking consecutive pairs of frames, resulting in a 240-dimensional vector. Additionally, the extracted features are combined with speaker-dependent vectors, called i-vectors~\cite{dehak2011language}, to form a 340-dimensional input used for neural network training. 

As highlighted in the main paper, the two network components of the RNN-T are responsible to carry out different tasks, hence different sSNU variants might be better suited for application in one or the other. Therefore, we investigated various configurations of our novel biologically-plausible variants.

\begin{table}[]
    \caption{RNN acronyms used in the result tables and details on the included parameters.}
    \label{tab:RNNsuffixSuppl}
    \centering
    \begin{tabular}{l|l|l|l|l|l}
         \hline
         RNN \textit{Suffix} & Comment & Thr. & \shortstack{Axo-\\dendritic} & \shortstack{Axo-\\somatic} & \multicolumn{1}{|c}{\shortstack{Axo-\\axonic}} \\
         \hline
         sSNU & Feedforward &  $\mathbf{b}$ & $\mathbf{W}$ &  \\
         sSNU \textit{R} & Recurrent &  $\mathbf{b}$ &  $\mathbf{W, H}$ & \\
         sSNU-a &  Adaptive thr. feedforward & $\mathbf{b_0}$ & $\mathbf{W}$ & & \\
         sSNU-a \textit{R} & Adaptive thr. recurrent & $\mathbf{b_0}$ & $\mathbf{W, H}$ & & \\
         sSNU-a \textit{Ra} & Adaptive thr. axo-somatic recurrent & $\mathbf{b_0}$ & $\mathbf{W, H}$ & $\mathbf{H_a}$ & \\
         sSNU-o &  Output modulating feedforward &  $\mathbf{b}$ & $\mathbf{W}$ & & $\mathbf{W_o, b_o}$ \\
         sSNU-o \textit{R} & Output modulating recurrent &   $\mathbf{b}$ & $\mathbf{W, H}$ & & $\mathbf{W_o, H_o, b_o}$ \\
         \hline
    \end{tabular}
\end{table}

The training is accomplished using the AdamW~\cite{Loshchilov2017Nov} optimizer with a one-cycle learning rate schedule~\cite{Smith2017Aug}, where the maximum learning rate $\eta_{m}$ has been determined for each run individually. We trained for 20 epochs wherein the learning rate was linearly ramped up from an initial value to a maximum value within the first six epochs and linearly ramped down to a minimum value in the subsequent 14 epochs, similar to~\cite{Saon2021Mar}. Table~\ref{tab:RNNsuffixSuppl} lists explicitly the trainable parameters for different sSNU variants along with their abbreviations. We trained with a batch size of $64$ on two V100-GPUs for approximately 10 days. To avoid overfitting we employed gradient clipping, in which the gradients are combined to a single vector $\boldsymbol{w}$ and the individual components are then computed as
\begin{align*}
\boldsymbol{\tilde{w}} = \boldsymbol{w} \odot \frac{c}{||\boldsymbol{w}||_2},
\end{align*}
with $c=1$ or $c=10$. In addition, we use dropout with a dropout probability of $p_W=0.25$ for all the input weights and $p_E=0.05$ for the embedding layer.

The final speech transcript was produced using beam search with a beam width of $16$. For the evaluation of our models, we followed the common procedure to report the word error rates (WER) on the Hub5 2000 Switchboard and the CallHome test set jointly~\cite{BibEntry2002Jan, Saon2021Mar}. As the baseline for our benchmark, we re-implemented the very recent state-of-the-art results from~\cite{Saon2021Mar}. We focused primarily on the dynamics of the neurons and thus used a basic model implementation without any external language model.
Such an LSTM-based RNN-T architecture achieves a WER of $12.7\,\%$, which we consider as our baseline.
However, it is worth mentioning that the research field of speech recognition is very active and although our selected baseline is representative of the state-of-the-art, it can potentially be improved with the features mentioned in~\cite{Saon2021Mar}.

As mentioned in the main paper, we followed a three-step approach to integrate our biologically-inspired units into the RNN-T network architecture. In addition to Section~4 of the main paper, the Tables~\ref{tab:hypPredNW},~\ref{tab:hypEncNW} and~\ref{tab:hypFullNW} contain more detailed hyperparameter settings of our simulations.

\begin{table}[h]
  \caption{Hyperparameters of the prediction network}
  \label{tab:hypPredNW}
  \centering
  \begin{tabular}{lcccc}
    \hline
    RNN & Cfg. & $\eta_m$ & $c$ & Additional\\
    \hline
    LSTM & 1x768 & $5 \cdot 10^{-4}$ & 10 &\\
    sSNU & 1x768 & $5 \cdot 10^{-4}$ & 10 & $d=0.9$ \\
    sSNU \textit{R} & 1x768 & $5 \cdot 10^{-4}$ & 10 & $d=0.9$\\
    sSNU-a  & 1x768 & $5 \cdot 10^{-4}$ & 10 & $d=0.9, \beta=0.1, \rho=0.9$\\ 
    sSNU-a \textit{R} & 1x768 & $5 \cdot 10^{-4}$ & 10 & $d=0.9, \beta=0.1, \rho=0.9$\\ 
    sSNU-o  & 1x768 & $5 \cdot 10^{-4}$ & 10 & $d=0.9$\\ 
    sSNU-o \textit{R} & 1x768 & $5 \cdot 10^{-4}$ & 10 & $d=0.9$\\ 
    \hline
    SNU & 1x768 & $5 \cdot 10^{-4}$ & 10 & $d=0.9$\\
    \hline
  \end{tabular}
\end{table}

\begin{table}[h]
  \caption{Hyperparameters of the encoding network}
  \label{tab:hypEncNW}
  \centering
  \begin{tabular}{lcccc}
    \hline
    RNN & Cfg. & $\eta_m$ & $c$ & Additional\\
    \hline
    LSTM & 6x640 & $5 \cdot 10^{-4}$ & 10 &\\
    sSNU-a \textit{Ra} & 6x640 & $5 \cdot 10^{-4}$ & 1 & $d=0.9, \beta=0.1, \rho=0.9$\\
    sSNU-o & 6x640 & $5 \cdot 10^{-4}$ & 10 & $d=0.9$\\
    sSNU-o \textit{R} & 6x640 & $9 \cdot 10^{-4}$ & 1 & $d=0.9$\\
    \hline
  \end{tabular}
\end{table}

\begin{table}[h]
  \caption{Hyperparameters of the full RNN-T network}
  \label{tab:hypFullNW}
  \centering
  \begin{tabular}{lcc|lcc|cc}
    \hline
    \multicolumn{3}{c}{Encoding network} & \multicolumn{3}{c}{Prediction network} & \multirow{2}{*}{$\eta_m$} & \multirow{2}{*}{$c$}\\
    RNN & Cfg. & Additional & RNN & Cfg. & Additional & & \\
    \hline
    LSTM & 1x768 & & LSTM & 6x640 & & $5 \cdot 10^{-4}$ & 10\\
    sSNU-o \textit{R} & 6x640 & $d=0.9$ & sSNU-a \textit{R} & 1x768 & $d=0.9, \beta=0.1, \rho=0.9$ & $9 \cdot 10^{-4}$ & 10\\
    sSNU-o \textit{R} & 1x768 & $d=0.9$ & sSNU-o \textit{R} & 6x640 & $d=0.9$ & $5 \cdot 10^{-4}$ & 1\\
    \hline
  \end{tabular}
\end{table}

\bibliography{bibliography}
\bibliographystyle{plainnat}


\section*{Acknowledgment}
We thank the Neuromorphic Computing and I/O Links group at IBM Research – Zurich for fruitful discussions and comments.


\end{document}